\documentclass[conference]{IEEEtran}

\usepackage{tabularx}
\usepackage[pdftex]{graphicx}
\usepackage{color}
\graphicspath{}
\DeclareGraphicsExtensions{.pdf,.jpeg,.png}

\begin{document}

\title{Optimizing Path ORAM for Cloud Storage Applications}


\author{\IEEEauthorblockN{Nathan Wolfe}
\IEEEauthorblockA{The Roxbury Latin School/MIT PRIMES\\ 
Email: nathanwolferl@gmail.com}
\and
\IEEEauthorblockN{Ethan Zou}
\IEEEauthorblockA{Lexington High School/MIT PRIMES\\ 
Email: zouethan19@gmail.com}
\and
\IEEEauthorblockN{Ling Ren}
\IEEEauthorblockA{MIT \\
Email: renling@mit.edu} \and
\IEEEauthorblockN{Xiangyao Yu}
\IEEEauthorblockA{MIT\\
Email: yxy@mit.edu}
}


%


\maketitle

\begin{abstract}
We live in a world where our personal data are both valuable and vulnerable to misappropriation through exploitation of security vulnerabilities in online services. For instance, Dropbox, a popular cloud storage tool, has certain security flaws that can be exploited to compromise a user’s data, one of which being that a user's access pattern is unprotected. We have thus created an implementation of Path Oblivious RAM (Path ORAM) for Dropbox users to obfuscate path access information to patch this vulnerability. This implementation differs significantly from the standard usage of Path ORAM, in that we introduce several innovations, including a dynamically growing and shrinking tree architecture, multi-block fetching, block packing and the possibility for multi-client use. Our optimizations together produce about a 77\% throughput increase and a 60\% reduction in necessary tree size; these numbers vary with file size distribution.

\end{abstract}


%
\IEEEpeerreviewmaketitle

\section{Introduction}
Nowadays, with many people possessing a multitude of devices, many of which have limited local storage, it is popular to store files in the cloud. Among the most popular cloud storage tools is Dropbox, which is compatible with computers and mobile devices, and allows users to store up to 2 GB of files for free (additional storage is available for a monthly fee).
For most users, Dropbox presents adequate security; all files are encrypted when stored on their servers. Nevertheless some will rightfully find certain issues with their system. One is that Dropbox can see which files you edit. This, known as the user’s access pattern, is a key piece of information, which Dropbox could potentially gather. Of course, Dropbox itself is trustworthy, but nowadays the US government has the power to request data from web services, and the access pattern may be included. While standard encryption offers some protection, it does nothing to hide a user’s access pattern.

Previous works have shown the importance in protecting access pattern. Memory access patterns can leak information such as a program’s control flow, which concerns conditional branches of the program and the order of commands executed \cite{HIDE}. Also, M.S. Islam et al. have shown how, from an encrypted email repository, adversaries can infer 80\% of search queries simply from access patterns \cite{Islam12}.

Oblivious RAM (ORAM) is a cryptographic primitive that completely eliminates the information leakage in memory access pattern.
In ORAM schemes, a client stores its data in encrypted and shuffled form on an untrusted server.
On each access, the client reads the untrusted memory, reshuffles that memory and possibly updates some state in trusted local storage called \emph{the client storage}.
Under ORAM, any memory access pattern is computationally indistinguishable from any other access pattern of the same length.

Since the initial ORAM proposal by Goldreich and Ostrovsky \cite{G87, O90, GO96}, researchers have striven to improve the efficiency and reduce the trusted storage requirement of ORAM \cite{OS97, WS08, BMP11, WS12, GM11, GMOT12, Kush12, SSS12, SCSL11, GENTRY13, PathORAM} for three decades. 
Of particular interest among them is Path ORAM \cite{PathORAM}.
Path ORAM’s bandwidth and client storage are both logarithmic in its capacity.
Furthermore, it is extremely simple and suitable for real world implementation. 
Immediately after its proposal in 2012, Path ORAM gained popularity in many applications due to its simplicity and efficiency. 

In this work, we adapted Path ORAM for use with oblivious cloud storage, and propose several optimizations specific to the cloud storage setting.
First, we have designed a system where ORAM on Dropbox can be used between multiple user computers. 
We also propose the optimizations of multi-block fetching, which improves performance by 51.9\%, and block packing, which saves 50-70\% storage overhead. 
Finally, we propose a dynamically growing and shrinking tree, providing the necessary storage size flexibility to work with Dropbox. All our optimizations together produce about a 77\% throughput increase, which varies with file size distribution.

Previous oblivious storage works include PrivateFS \cite{PrivateFS}, Shroud \cite{Shroud} and ObliviStore \cite{oblivistore}.
ObliviStore, which is based on the SSS ORAM \cite{SSS12} with careful engineering claims the best memory throughput.
A recent work named Burst ORAM \cite{BurstORAM}, built on top of ObliviStore, optimizes response time instead of throughput.
We remark that the goal of this paper is not to claim superiority of Path ORAM over ObliviStore or Burst ORAM.
Rather, the ideas in this paper apply to those constructions as well, and we simply choose Path ORAM to evaluate our proposal due to its conceptual simplicity. 

\textbf{Organization.} 
We will first provide some background on Path ORAM in Section \ref{sec:Background}. We will then explain the system architecture behind our ORAM implementation in Section \ref{sec:SysArch}. Then, after explaining our various optimizations and improvements in Section \ref{sec:Opt}, we will present some evaluations results in Section \ref{sec:Eval}.

\section{Background on Path ORAM}

\label{sec:Background}
\begin{figure}
\begin{center}
  \includegraphics[width=6cm]{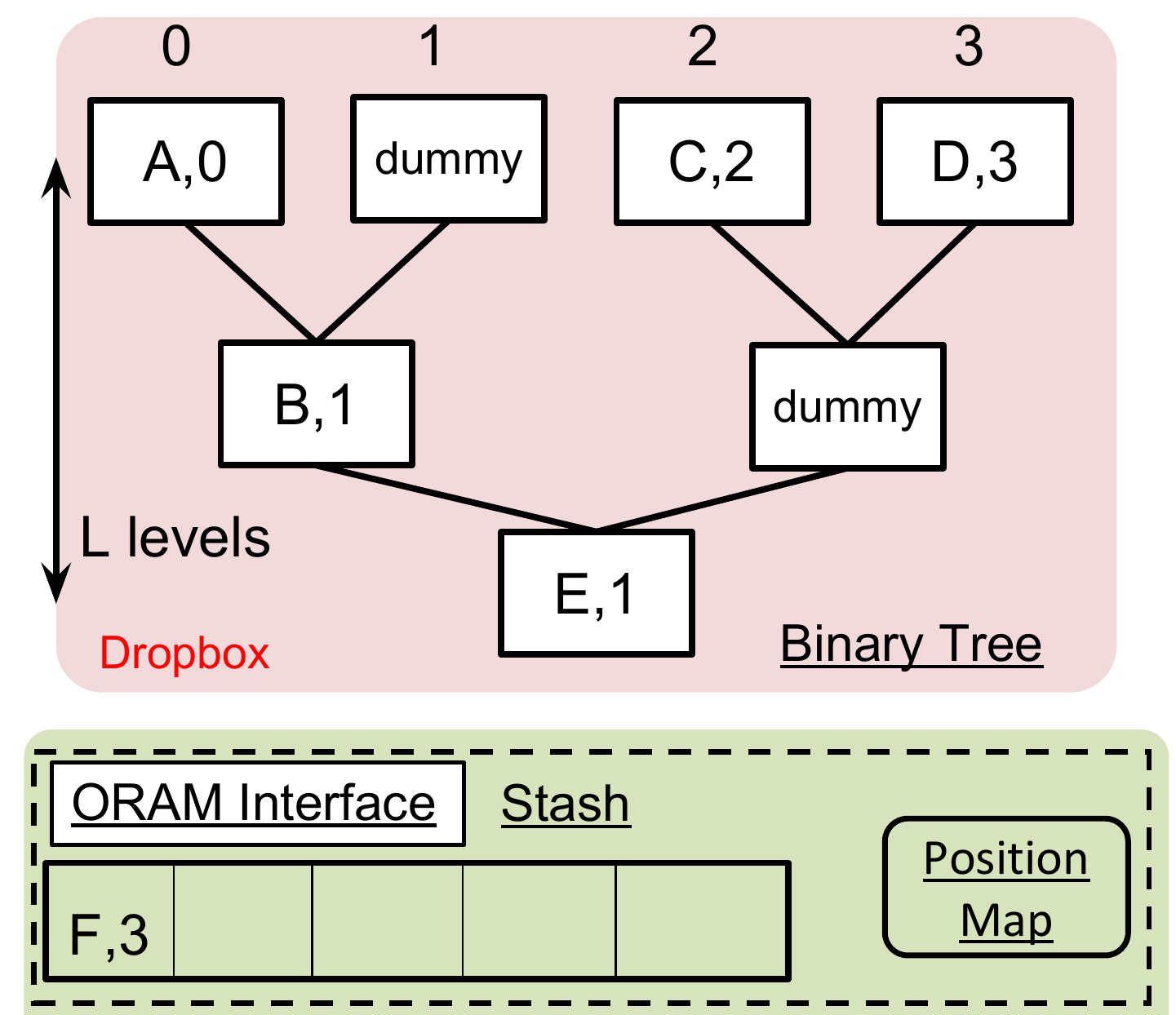}
  \caption{An example of a Path ORAM construction consisting of the binary tree (untrusted) and the stash and position map (trusted).}
  \label{fig:bintree}
\end{center}
\end{figure}

In Path ORAM, data blocks are arranged into a binary tree, with up to a constant number blocks at each node of the tree (Figure \ref{fig:bintree}). 
Each block is mapped to random path of the tree.
To access a block, the Path ORAM algorithm accesses all blocks along the target’s path. 

We summarize how Path ORAM works in Figure \ref{fig:bintree}. The reader can refer to \cite{PathORAM} for more details. The binary tree is stored on the server. 
Suppose it has height $L$ and $2^L$ leaves. 
The levels are numbered 0 to $L$, where level 0 is the root and level $L$ contains the leaves (in a complete tree). 
Each node in the binary tree is called a bucket, and contains $Z$ blocks. 
If a bucket contains less than $Z$ “real” blocks, the remainder are composed of dummy blocks, which are indistinguishable from real blocks after encryption.

On the client side, we construct a data structure called the {\bf stash}, which stores a small subset of the data blocks, and a {\bf position map} that assigns each block to a path in the tree. Every path is uniquely defined as leading from the root to one of the $2^L$ leaf buckets. Importantly, mapping is random, allowing for multiple unrelated blocks to be mapped to one leaf. This is shown in Figure \ref{fig:bintree}: for example, both blocks B and E are mapped to leaf 1, and are required to reside on the path defined by this leaf (referred to as path 1). The position map is updated as blocks are accessed and their leaf assignments changed. This is critical because Path ORAM maintains the invariant that if the block is not on its assigned path, it is in the stash (if it is not, there is a bug).

Reading/writing a block consists of several steps. First, we consult the position map to determine what path the target block is on. Then, we fetch the blocks on that path from the tree. Based on the invariant just mentioned, it should be among those fetched blocks or in the stash. We can then securely access the target block. After making any necessary changes to the block we assign it to a different path using the position map. Finally we take all blocks originally fetched and write them back into the tree.

In the past many papers have focused on Path ORAM systems and their optimizations. However, this is the first applying Path ORAM directly for use with Dropbox, and the first to explore optimizations unique to the Dropbox usage case.

\section{System Architecture}

\label{sec:SysArch}
We introduce the general organization of our design and several changes to the basic implementation to better suit Dropbox.

\subsection{Big Picture}

\begin{figure}
\begin{center}
  \noindent\includegraphics[width=8.5cm]{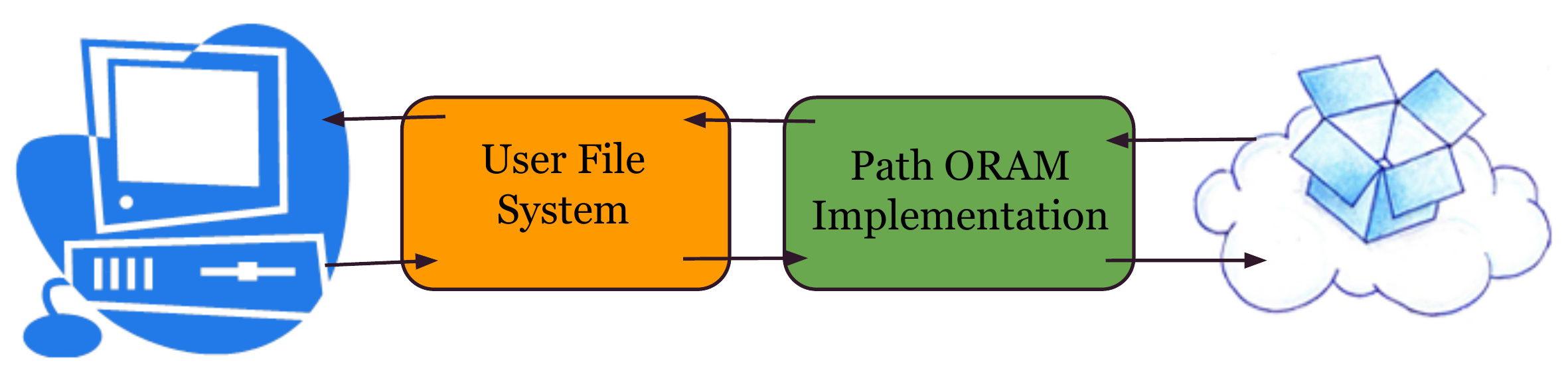}
  \caption{Basic structure of our Path ORAM implementation}
  \label{fig:metadiag}
\end{center}
\end{figure}

The overall interface of our design consists of three parts: the client, the Path ORAM implementation and Dropbox. Instead of the client directly writing and reading files to the Dropbox folder, we implemented the ORAM as an intermediary controller between the two. So a user will directly interact with the User File System (See Section \ref{sec:UFS}), which in turn interacts with the ORAM controller to read and write files. The ORAM controller interacts with Dropbox by having the buckets of the tree written to the Dropbox folder. The Dropbox service will then sync the contents to the cloud. When reading a file, the bucket files are downloaded. As we will discuss in Section \ref{sec:MultiComp}, a system with multiple computers accessing the same files requires the stash, position map, and other dictionaries to be written to Dropbox as well, which can be downloaded by other computers in the system. These structures which are normally kept in secure storage have to be encrypted with a secret key.

\subsection{User File System}

\label{sec:UFS}
Dropbox's baseline functions allow users to read and write files to a remote server in the cloud. Thus, we needed a way for our ORAM system to support the reading and writing of files of various sizes. Our solution was implementing the User File System, which enables the writing of files of different sizes. 

The User File System interacts with the ORAM controller and provides the users with an interface to read, update and delete files of arbitrary size. The User File System consists of several dictionaries which hold vital metadata. Its main feature is to partition files into chunks, whose size is a parameter we call {\bf segment size}. These chunks are then each written to or read from the ORAM.

\begin{figure}
\begin{center}
\includegraphics[width=8cm]{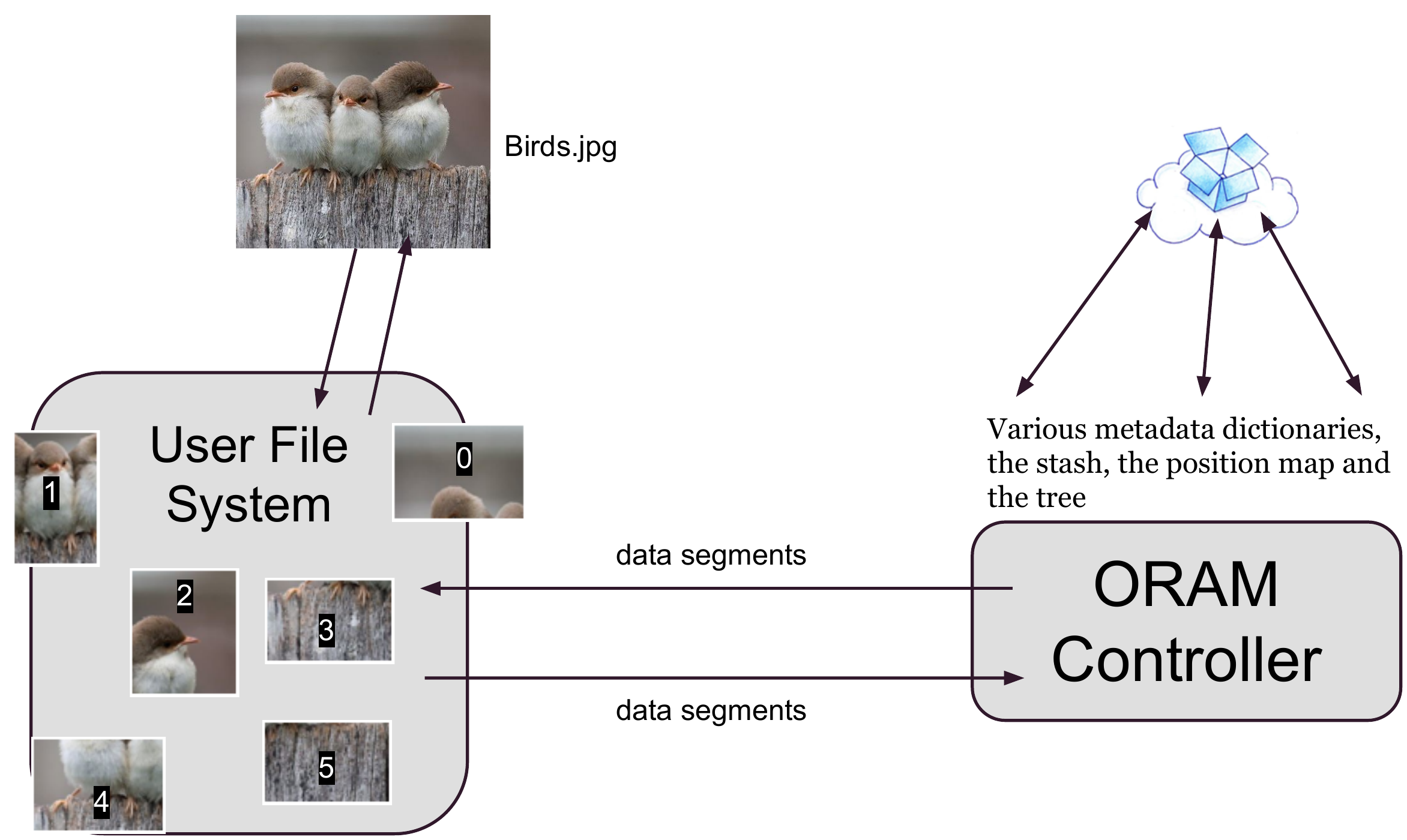}
\caption{Visual representation of how the User File System splits and then reads/writes files from/to Dropbox.}
\label{fig:userfilesys}
\end{center}
\end{figure}

To write a file, we split it up into subfiles each containing a number of bytes equal to (or less than, in the case of the last subfile) the segment size. We then assign each subfile a unique segment ID (stored in a dictionary) that allows us to be able to track the individual segments when we want to read the file again. Then, we write each file segment to the ORAM and update another dictionary with the number of partitioned pieces in that particular file. When reading a file, we look up the file name in a dictionary to retrieve the number of segments in that file. Then, we obtain the individual segment IDs of each data segment and fetch them from the ORAM. The data is then stitched together and returned to the user. Deleting a file works similarly to reading. After looking up the segment IDs of the file fragments, we simply delete those data from the ORAM and delete their corresponding information from the two metadata dictionaries.

\subsection{Multi-Computer}

\label{sec:MultiComp}
One of the major reasons cloud storage programs such as Dropbox are so popular is that a user can access and change files on one computer, and then move to another computer to access the same files. In order to support this feature, we added functionalities to the file systems.

When a user “logs off” from the computer, the dictionaries in the User File System, the position map, and the stash are written to the Dropbox folder, and then synced to the remote server. An important implication is that these files are encrypted and padded to a constant file size when written to Dropbox to ensure security and prevent an adversary from gaining information from file size or access pattern. When the user wants to access his files from another computer, he/she downloads these files from the server to reconstruct the User File System and ORAM metadata structures.

\subsection{Initial Evaluations}

Our current implementation is significantly slower than directly using Dropbox without ORAM. Thus, the implementation has some apparent drawbacks. The major problem comes from partitioning an immensely large file into millions of segments (based on the assigned segment size), which means that reading or writing one large file takes a very long time. Along with this, a static tree size is a problem when using Dropbox because we don’t want space to be wasted (more space equals more money). With these issues in mind, we sought to develop optimizations to improve upon our initial implementation.

\section{Optimizations}

\label{sec:Opt}

In this section we provide detailed explanations and motivation for the new optimizations we have implemented: multi-block fetching, block packing and a dynamically growing and shrinking tree size.

\subsection{Multi-Block Fetching}

One idea to improve performance dramatically is to fetch multiple data blocks in each access, thus cutting down on the total number of accesses required. Unfortunately, usually when using ORAM, it is unclear how data blocks are grouped outside, so it is hard to know which groups of blocks to fetch together. However, in our implementation of ORAM with Dropbox, the data segments stored can be grouped in a defined way, with each grouping corresponding to a particular file. Thus we can cut down on the number of tree accesses we perform by putting segments from the same file on the same path in the tree and fetching them at the same time. The technique is as simple as grouping segments from each file into n-tuples, mapping them all to the same path, and then fetching them all at the same time. Theoretically this should reduce the number of accesses needed by a factor of n. Unfortunately, however, grouping a great number of segments together in the same path produces inefficiency. Any individual path may become congested, slowing down the operation. Through evaluations of file access speed we have managed to find an optimal value for n for this grouping process.

\subsection{Block Packing}

\label{sec:mult-blkOpt}
One way to speed up the ORAM operations is to cut down on the number of partitions of a file. In order to do that, we need to use a large segment size. However, larger segment sizes hurt small files because a tremendous amount of space is wasted, and it takes longer to read small files. 

With this in mind, our approach to improving this aspect of the ORAM was to implement block packing, which essentially packs more than one file segment together into one block. The motivation behind this approach comes from the idea that a small file does not necessarily need to be the only file present in a block. Additionally, leftover data that does not divide evenly into the segment size can also be packed into the empty spaces of a block. By doing this, space is saved and not wasted, resulting in a smaller tree and smaller path lengths (see Section \ref{sec:growshrink}), and consequently reduces file access time (though only slightly).

Block packing works as follows. We add two more dictionaries to the User File System: one holds the amount of space left in a block and the other holds the start and end offsets of a certain file segment inside of a block. We write a file of an arbitrary size. The “full” segments of data (take up an entire block) are written normally to the ORAM. We then look for available space in a block for the very last segment (which, in the case of a small file will be the entire file). If we find a block that already has some data in it and that has sufficient excess space, we append the final segment to the end of the already existing data, and update the dictionaries accordingly. If we do not find an available block, we write the data to an empty block and add the ID of that block to the dictionary with its amount of empty space. When we want to read the file, all the segments except the last are read normally. We find the position of the last segment in the block and read out the specific portion of the data.

Deleting a file becomes slightly more complicated. All file segments except the last one are deleted normally. When deleting the last segment of a file, we run into the problem of there possibly being a “hole” of no data inside a block. Our solution was to shift the data after the deleted segment as far up in the block as possible so that there are no gaps in the middle of a block. We then update the dictionaries for the shifted segments of data.

\subsection{Dynamic Growing and Shrinking Tree}

\label{sec:growshrink}
A normal ORAM implementation involves a storage space that is fixed in size (e.g., main memory or hard drive of a computer). However, Dropbox presents the unique problem of a dynamically sized storage space. First, users may wish to store sensitive files in an ORAM system alongside less important files stored unencrypted in Dropbox. By supporting resizing of the tree we support this usage scenario. Second, Dropbox allows users to increase the size of their storage by either paying or sharing the service with friends. By supporting tree resizing we allow users to change their storage space without having to create a new ORAM library each time. Thus, resizing provides convenience to the user in a way unique to this project.

The first problem we encounter is to determine when to resize. Previous work suggested that the optimal utilization (meaning the fraction data block slots that are used) is around 50\% \cite{oram-isca13}.

Our implementation supports both growing and shrinking. Both functions are triggered based on current utilization,. When the utilization goes above a certain threshold we grow the tree until the utilization goes back down to a target level. When it goes below a certain threshold we shrink (to save Dropbox space), with the utilization going back up to a target level.

Growing the tree works as follows. First, we add the correct number of buckets to the leaf end of the tree. Obviously this reveals the information that we are growing, and that the total size of our files stored has gone beyond some threshold, but that information is both non-critical and practically unavoidable to leak. We then correct all leaves in both the position map and the stash in order to conform to the new leaf numbers. Nodes with new children are invalidated as leaves, so records pointing to these nodes are reassigned to one of the children.

Shrinking works similarly; first we remove the correct number of buckets from the leaf end of the tree, dumping their contents into the stash. Again, this leaks the information that our total file size must have decreased. We then correct all leaves in the position map and the stash by truncating the path so that it lines up with the correct leaf, and leave the ones in the tree to sort out later.

\section{Evaluations}

\label{sec:Eval}
In this section, we evaluate the performance of our optimizations. 
\subsection{Methodology}

The CPU Model of the machine used for the experiments is Intel(R) Core\textsuperscript{TM} i3-2100 CPU @ 3.10GHz. The disk model is ST31000524AS (7200 RPM). Windows 7 operating system and Python 3.4 were used. Our encryption algorithm was Advanced Encryption Standard (AES) 128-bit in counter (CTR) mode, in the PyCrypto package v2.6.1.

\begin{figure}
\begin{center}
\noindent\includegraphics[width=8.5cm]{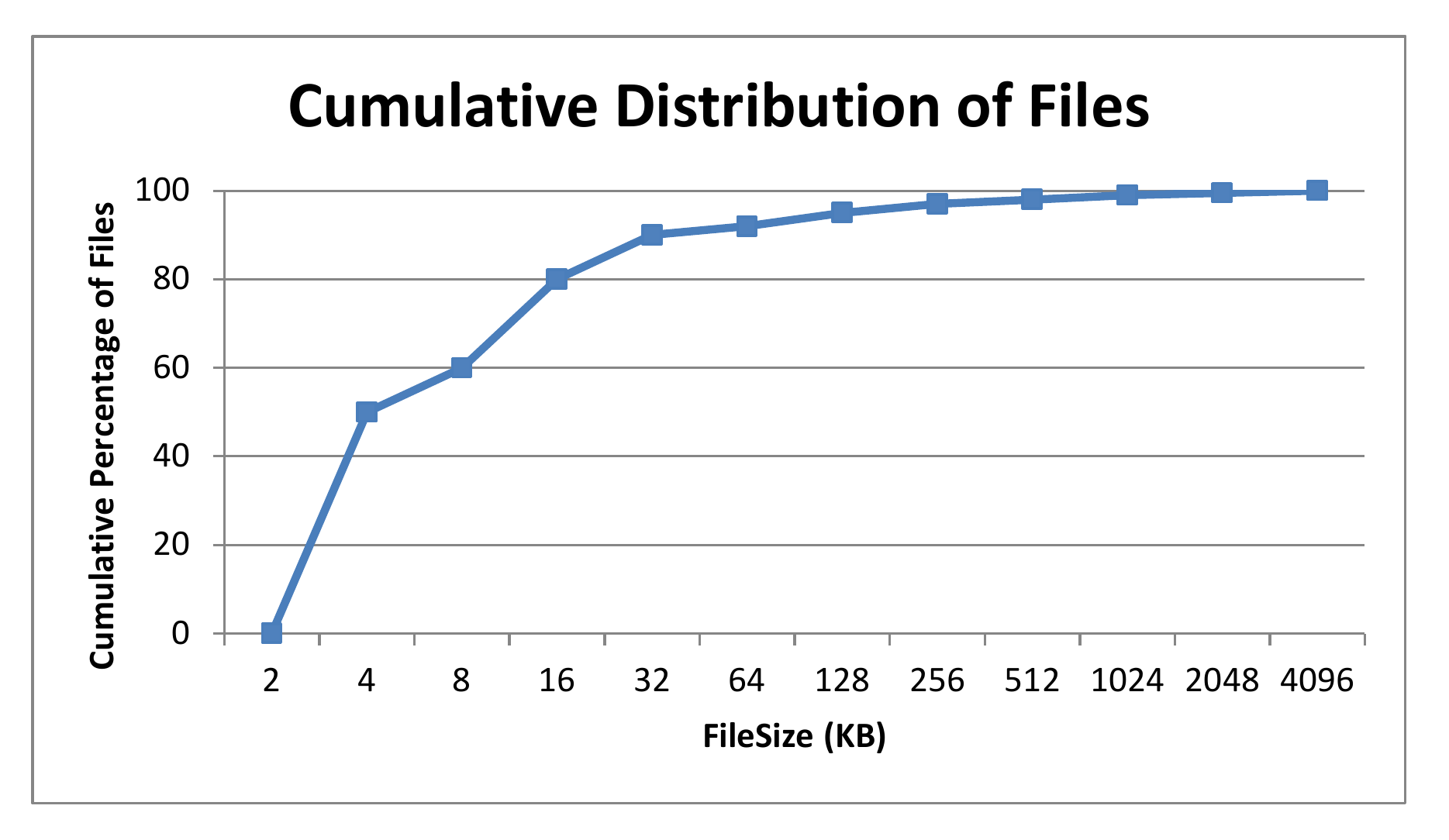}
\caption{File Distribution Function showing the cumulative percentage of files of various sizes (KB).}
\label{fig:distfun}
\end{center}
\end{figure}

For all the experiments except the one in Section \ref{sec:BlkPack}, we utilize a distribution of files of various sizes to better simulate the contents/actions of a “real” file system, by reading/writing 1000 files based upon this cumulative distribution. This file distribution was chosen from previous work done on file system workloads \cite{roselli2000comparison}, and is reproduced in Figure \ref{fig:distfun}.

For all experiments, the parameters were kept at constant values (unless we were testing that parameter). 
The ORAM tree size (if auto resizing is off) was dependent on the amount of data written to it, twice the total file size. 
Segment size was 64 KB (optimal as indicated by Figure \ref{fig:segsize}) with bucket sizes of $Z$ = 3.  
The maximum stash size was 100. 
The utilization ratio that triggers tree growth was 0.45 and for tree shrinkage 0.55. 
We use throughput in Mega Byte per second (MB/sec) as the performance metric.

\subsection{Optimal Segment Size}
\begin{figure}
\begin{center}
\noindent\includegraphics[width=8.5cm]{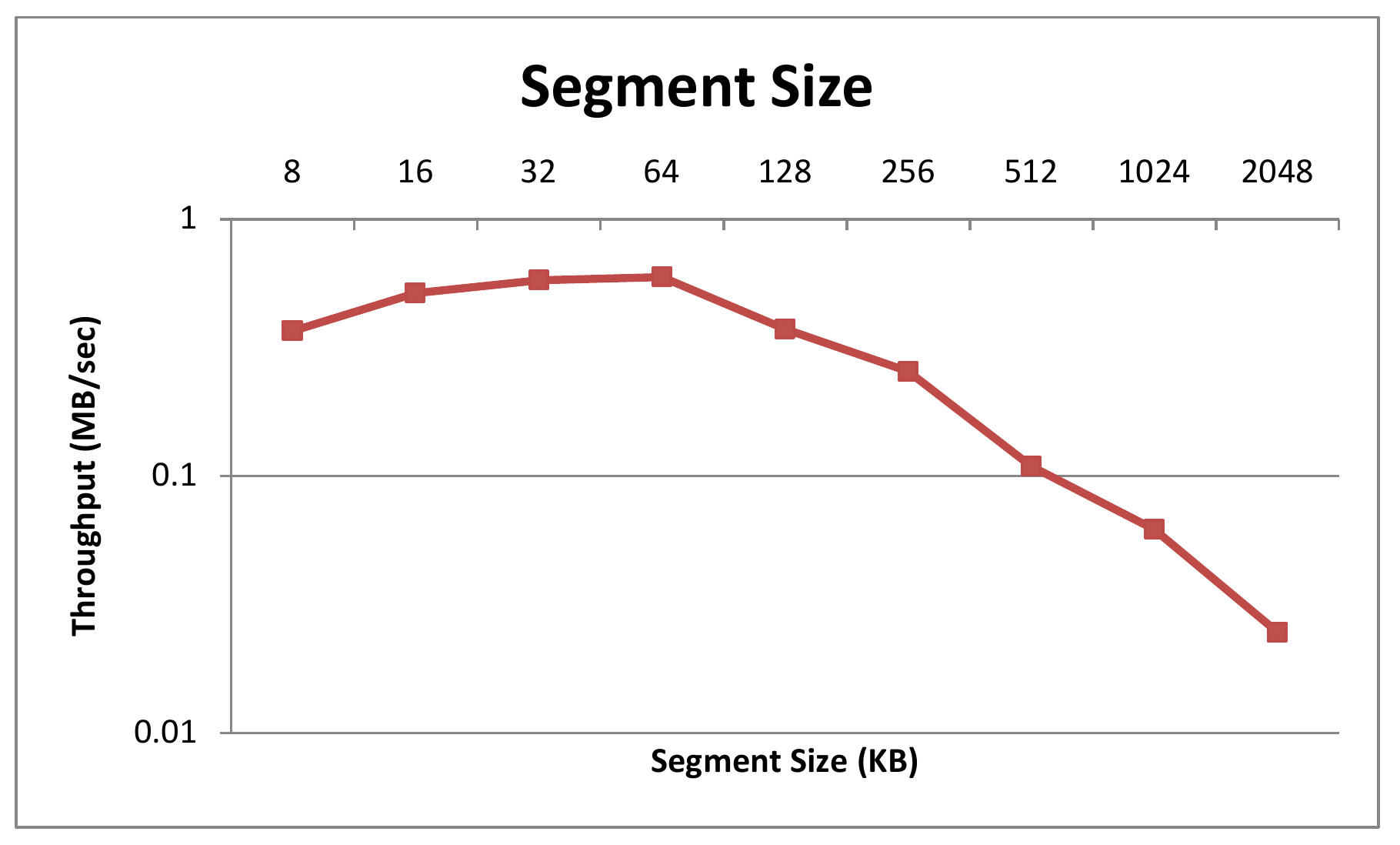}
\caption{Sweep study comparing segment size with throughput (in MB/sec)}
\label{fig:segsize}
\end{center}
\end{figure}

We performed a sweep study to evaluate the optimal segment size into which files are partitioned for storage. The sweep study in Figure \ref{fig:segsize} shows that optimal segment size is 64 KB. It is not surprising that there is an optimal segment size. For small segment sizes, large files are handled inefficiently because they gets partitioned into many more pieces, resulting in a massive number of operations to read/write one file. For large segment sizes, small files take longer to read because of the majority of filler data in a block.

\subsection{Initial ORAM Efficiency}
\newcolumntype{C}[1]{>{\centering\let\newline\\\arraybackslash}p{#1}}
\begin{table}
\begin{center}
  \begin{tabular}{| C{1.5cm} | C{3cm} | C{2.5cm} |}
  \hline
  \bf{Disk} & \bf{Disk + Encryption} & \bf{Disk + ORAM} \\
  \hline
  42.13 & 24.42 & 0.66 \\
  \hline
\end{tabular}
\vspace{6pt}
\caption{Throughput (in MB/sec) of Path ORAM (before our optimization) and comparisons to two baselines.}
\label{tab:baseline}
\end{center}
\end{table}

We evaluated the overhead cost of implementing of Path ORAM with that of encryption alone. As a baseline we compared this to reading/writing files to the local disk (Shown in Figure \ref{tab:baseline}).  Encryption added a little overhead (about 1.7 times slower) to the operations. Incorporating ORAM made the operations an additional 37 times slower when compared to encryption alone. As indicated by the table, the performance of our Dropbox-ORAM implementation without any optimizations is very slow, as expected. The remainder of this section evaluates the effectiveness of our improvements.

\subsection{Multi-Block Fetching}
\begin{figure}
\begin{center}
\noindent\includegraphics[width=8.5cm]{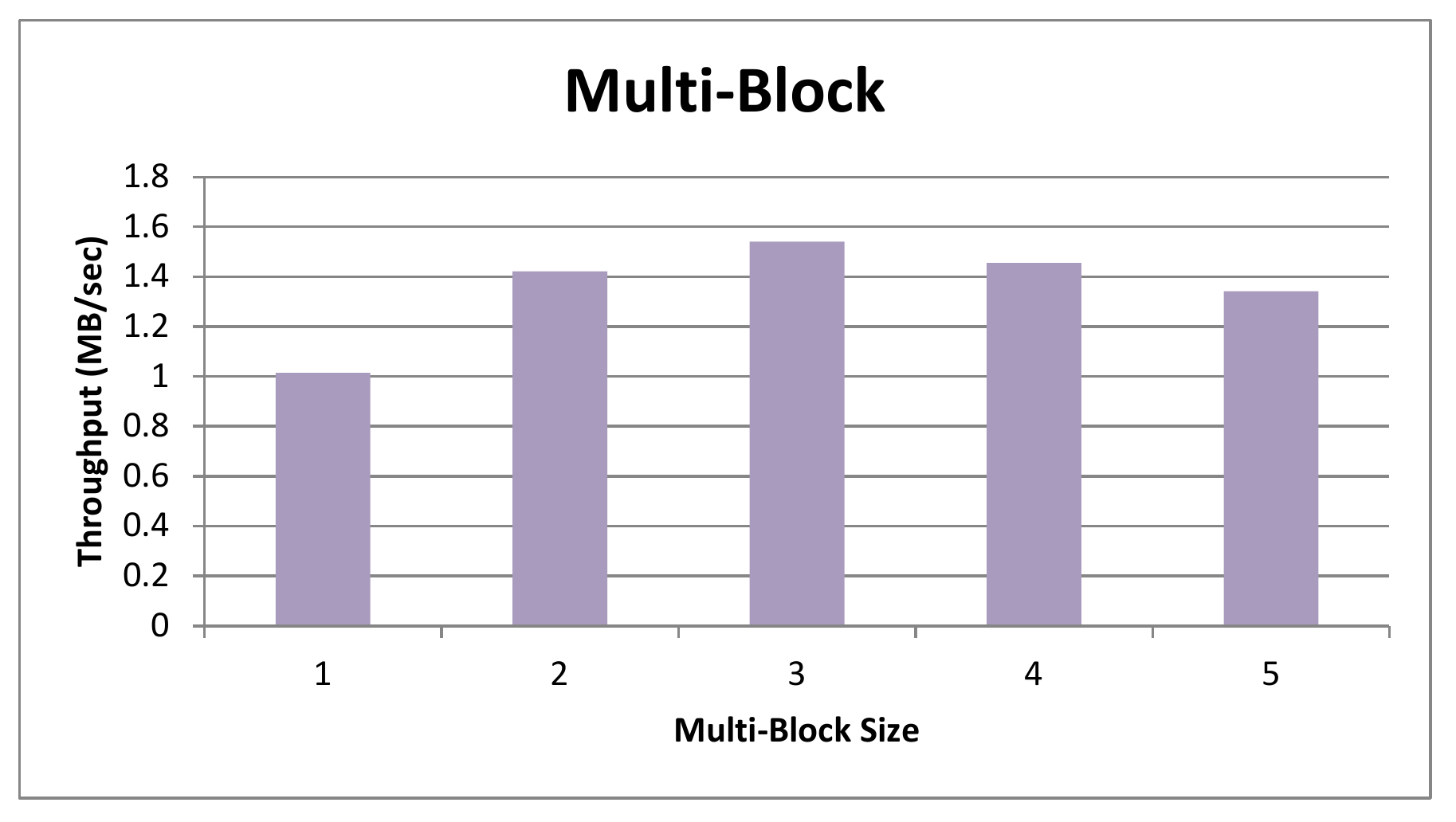}
\caption{Throughput (in MB/sec) of Path ORAM at different Multi-Block sizes.}
\label{fig:multiblock}
\end{center}
\end{figure}

As noted in Section \ref{sec:mult-blkOpt}, we utilized multi-block fetching with segment grouping to improve the efficiency of our tree architecture. From Figure \ref{fig:multiblock} we can see that the optimal grouping size appears to be 3, which also indicates that multiple block fetching does improve performance. However, larger grouping sizes create more congestion in some paths. To compensate for congestion, a technique called background eviction has to be employed, which takes additional overhead \cite{oram-isca13, Yu14DSB}.

\subsection{Block Packing}

\label{sec:BlkPack}
\begin{figure}
\begin{center}
\noindent\includegraphics[width=4.3cm]{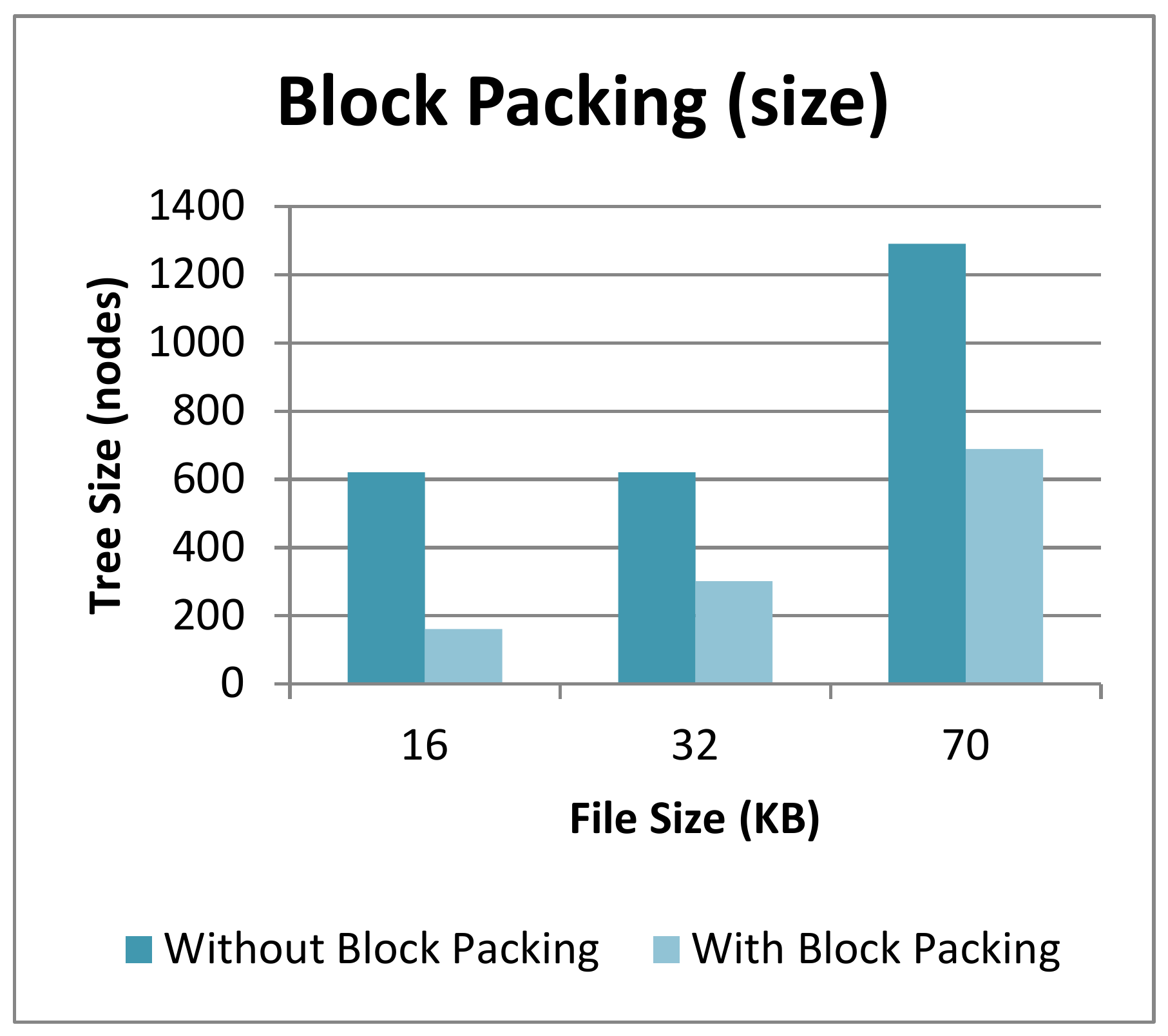}\includegraphics[width=4.3cm]{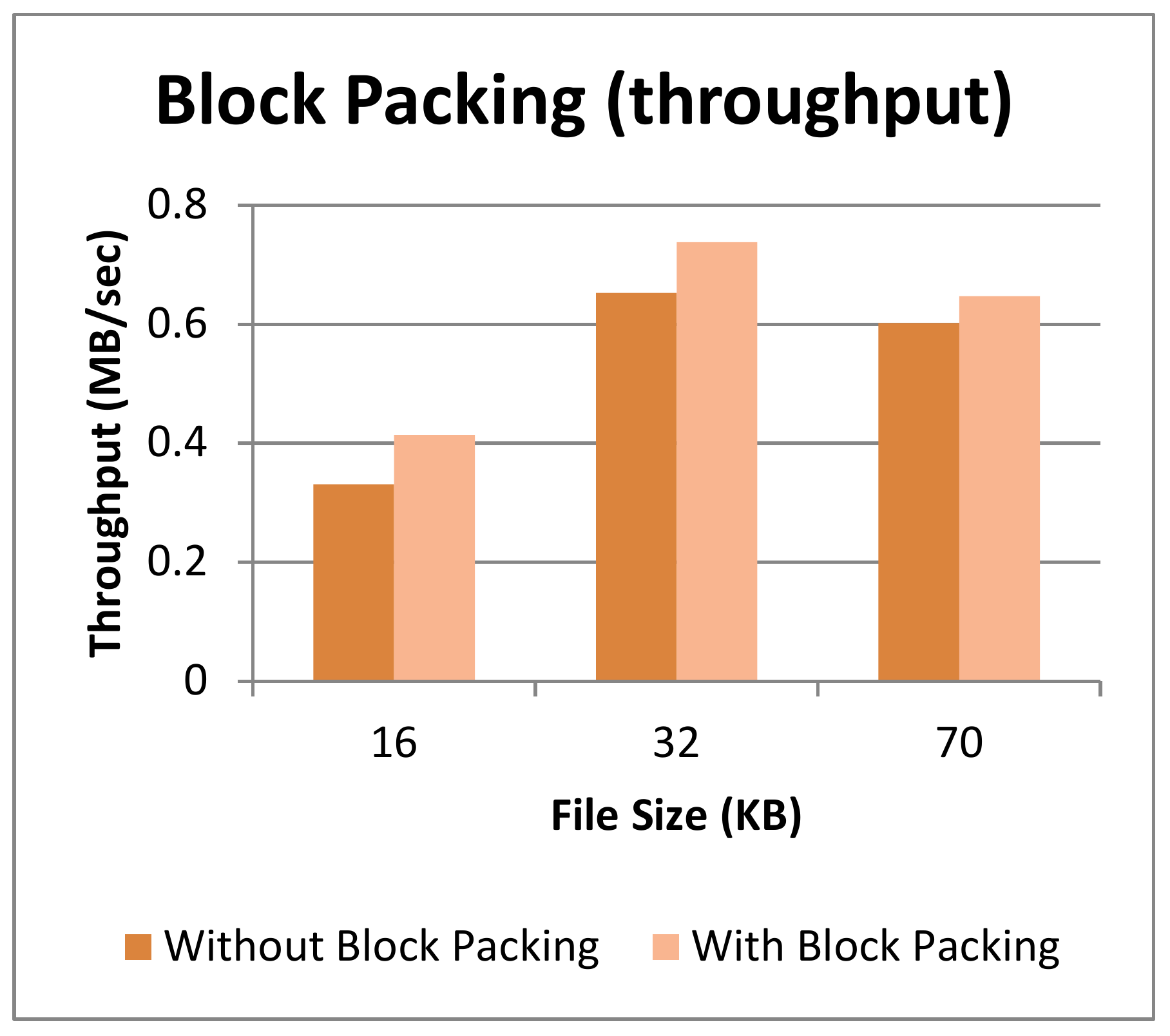}
\caption{Block Packing evaluations comparing tree size (in nodes) and throughput (in MB/sec) with various file sizes. }
\label{fig:blockpack}
\end{center}
\end{figure}

To evaluate the impact of block packing on performance, we assessed two dependent parameters, tree size and data throughput. For both tests, we wrote 1000 files of the same size to the ORAM and then recorded the throughput of reading the files 1000 times. Automatic resizing was turned on in order to illustrate the space saving aspect of this optimization. The file sizes of 16, 32 and 70 KB were chosen specifically for this test. Without block packing, the size of the tree was expected to be the same for both the 16 KB files and the 32 KB files because each file will take up one block, and with block packing, the size was expected to be $1/4$ and $1/2$ of the previous sizes because 4 times and 2 times as many files can be packed together in a block, respectively. The results in Figure \ref{fig:blockpack} confirm our expectations. The 70 KB file was chosen to illustrate that block packing works with the leftover segments. As shown in the graph, the tree was about two times smaller, because the extra 6 KB file segments were packed together, saving space. Block packing also yielded a small improvement in data throughput. The saved space results in a smaller tree structure. As a result the time taken for a path read/write is expected to be somewhat shorter.

We tested this optimization only in tandem with dynamic growing and shrinking. The reason for this was that the main goal of block packing is to save space, and so the only way a user will benefit is if the tree shrinks to a smaller size accordingly. The reduced tree size made possible by block packing then provides a significant performance increase, seen in Figure \ref{fig:blockpack}. Without dynamic growing and shrinking, block packing produces a negligible performance increase.

\subsection{Dynamic Growing and Shrinking}
\begin{figure}
\begin{center}
\noindent\includegraphics[width=8.5cm]{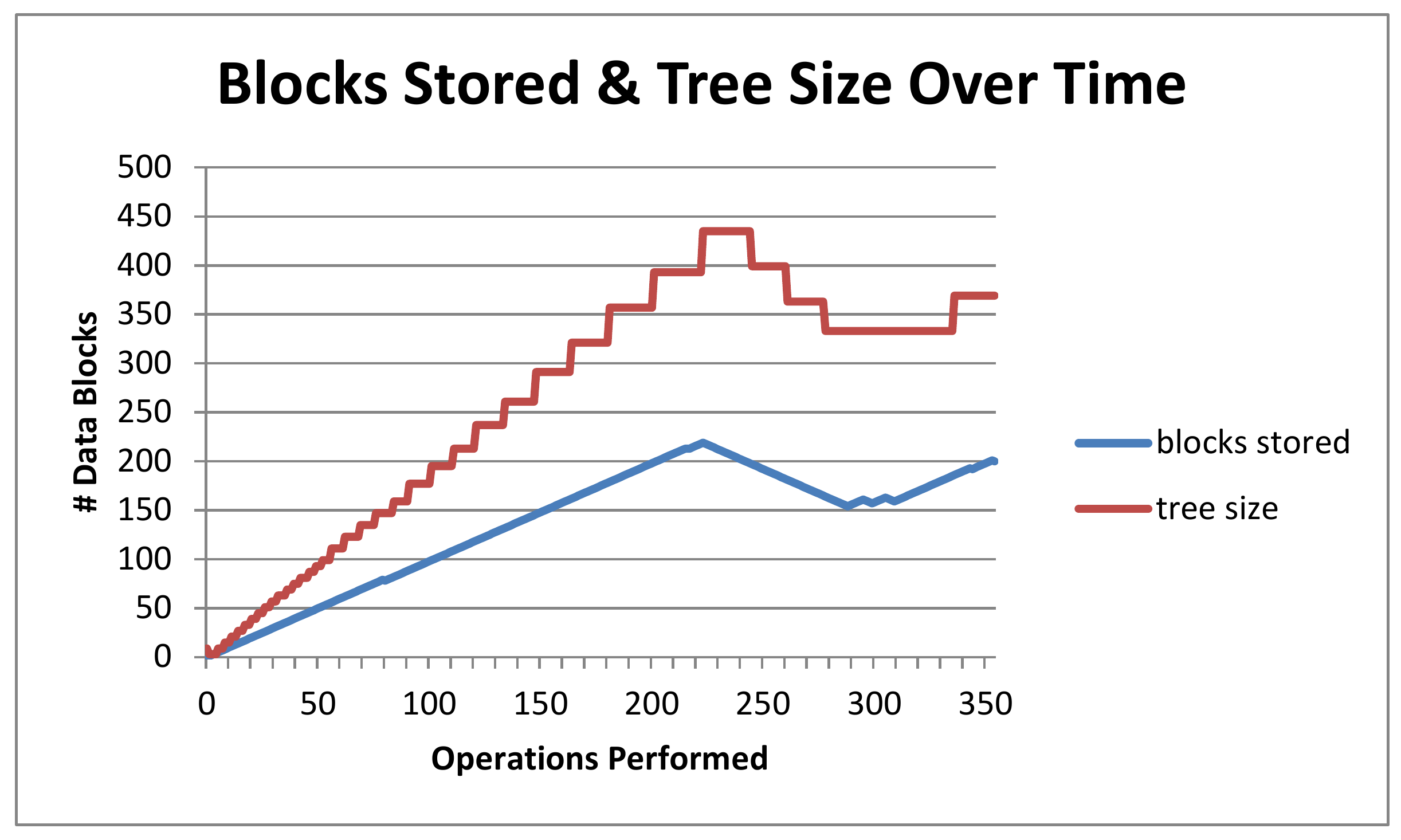}
\caption{Tree resizing demonstration charting the number of ``real blocks'' stored in the tree and the size of tree over time.}
\label{fig:resizing}
\end{center}
\end{figure}

To demonstrate the implementation of our dynamic tree structure, we tested the impact of random file addition or deletion to the tree structure. Thirty-two actions were performed using files of random sizes for this demonstration. Figure \ref{fig:resizing} displays the impact of an increasing and decreasing total file size stored in the tree. Whenever the tree utilization breaks the threshold set at $0.45$ or $0.55$, the tree to resizes to fulfill the target 50\% utilization.

\subsection{Total Optimization Impact}

All of our improvements taken together yield significant savings in both storage space and time. Block packing alone yielded a performance gain on average of 15\%, and multi-block fetching with groups of 3 yielded a performance gain of 52\%. With both optimizations running together we would expect total gain of about 60\%. The estimated throughput is 1.06 MB/s. 

\section{Conclusion}
Using an ORAM system with Dropbox allows the user to hide a key piece of information, his/her access pattern. Doing so, however, costs a significant overhead. To counteract this, we have introduced a few optimizations, which, as shown in the evaluations section, cut down on the overhead and increase throughput. Additionally, to conform best with Dropbox, we have altered the Path ORAM system with improvements, including potential for multi-computer use and a dynamically resizing tree. These developments are unique to any Path ORAM work that has been previously described. Our goal is that continued optimization will increase Path ORAM’s speed sufficiently that it may be practical for mainstream use. 

To summarize our work, we first designed and implemented a Path ORAM system to work with Dropbox. We then performed optimizations including multi-block fetching, block packing, and a growing and shrinking tree, improving performance by an estimated 77\%, and reducing storage overhead by 50-70\%. Both of these savings provide us with flexibility that is necessary for Path ORAM to work harmoniously with Dropbox.

In the future we plan to make our work into a public full-fledged software package, with a crash recovery mechanism and possibly other optimizations.




%



\bibliographystyle{IEEEtran}
\bibliography{paper}

\begin{thebibliography}{10}
\providecommand{\url}[1]{#1}
\csname url@samestyle\endcsname
\providecommand{\newblock}{\relax}
\providecommand{\bibinfo}[2]{#2}
\providecommand{\BIBentrySTDinterwordspacing}{\spaceskip=0pt\relax}
\providecommand{\BIBentryALTinterwordstretchfactor}{4}
\providecommand{\BIBentryALTinterwordspacing}{\spaceskip=\fontdimen2\font plus
\BIBentryALTinterwordstretchfactor\fontdimen3\font minus
  \fontdimen4\font\relax}
\providecommand{\BIBforeignlanguage}[2]{{%
\expandafter\ifx\csname l@#1\endcsname\relax
\typeout{** WARNING: IEEEtran.bst: No hyphenation pattern has been}%
\typeout{** loaded for the language `#1'. Using the pattern for}%
\typeout{** the default language instead.}%
\else
\language=\csname l@#1\endcsname
\fi
#2}}
\providecommand{\BIBdecl}{\relax}
\BIBdecl

\bibitem{HIDE}
\BIBentryALTinterwordspacing
X.~Zhuang, T.~Zhang, and S.~Pande, ``Hide: An infrastructure for efficiently
  protecting information leakage on the address bus,'' in \emph{Proceedings of
  the 11th International Conference on Architectural Support for Programming
  Languages and Operating Systems}, ser. ASPLOS XI.\hskip 1em plus 0.5em minus
  0.4em\relax New York, NY, USA: ACM, 2004, pp. 72--84. [Online]. Available:
  \url{http://doi.acm.org/10.1145/1024393.1024403}
\BIBentrySTDinterwordspacing

\bibitem{Islam12}
M.~S. Islam, M.~Kuzu, and M.~Kantarcioglu, ``Access pattern disclosure on
  searchable encryption: Ramification, attack and mitigation.'' in \emph{NDSS},
  2012.

\bibitem{G87}
\BIBentryALTinterwordspacing
O.~Goldreich, ``Towards a theory of software protection and simulation by
  oblivious rams,'' in \emph{Proceedings of the Nineteenth Annual ACM Symposium
  on Theory of Computing}, ser. STOC '87.\hskip 1em plus 0.5em minus
  0.4em\relax New York, NY, USA: ACM, 1987, pp. 182--194. [Online]. Available:
  \url{http://doi.acm.org/10.1145/28395.28416}
\BIBentrySTDinterwordspacing

\bibitem{O90}
R.~Ostrovsky, ``Efficient computation on oblivious rams,'' in \emph{Proceedings
  of the Twenty-second Annual ACM Symposium on Theory of Computing}, ser. STOC
  '90.\hskip 1em plus 0.5em minus 0.4em\relax New York, NY, USA: ACM, 1990, pp.
  514--523.

\bibitem{GO96}
O.~Goldreich and R.~Ostrovsky, ``Software protection and simulation on
  oblivious rams,'' \emph{J. ACM}, vol.~43, no.~3, pp. 431--473, May 1996.

\bibitem{OS97}
\BIBentryALTinterwordspacing
R.~Ostrovsky and V.~Shoup, ``Private information storage (extended abstract),''
  in \emph{Proceedings of the Twenty-ninth Annual ACM Symposium on Theory of
  Computing}, ser. STOC '97.\hskip 1em plus 0.5em minus 0.4em\relax New York,
  NY, USA: ACM, 1997, pp. 294--303. [Online]. Available:
  \url{http://doi.acm.org/10.1145/258533.258606}
\BIBentrySTDinterwordspacing

\bibitem{WS08}
P.~Williams and R.~Sion, ``Usable pir.'' in \emph{NDSS}, 2008.

\bibitem{BMP11}
D.~Boneh, D.~Mazieres, and R.~A. Popa, ``Remote oblivious storage: Making
  oblivious ram practical,'' 2011.

\bibitem{WS12}
P.~Williams and R.~Sion, ``Single round access privacy on outsourced storage,''
  in \emph{Proceedings of the 2012 ACM Conference on Computer and
  Communications Security}, ser. CCS '12.\hskip 1em plus 0.5em minus
  0.4em\relax New York, NY, USA: ACM, 2012, pp. 293--304.

\bibitem{GM11}
M.~T. Goodrich and M.~Mitzenmacher, ``Privacy-preserving access of outsourced
  data via oblivious ram simulation,'' in \emph{Proceedings of the 38th
  International Conference on Automata, Languages and Programming - Volume Part
  II}, ser. ICALP'11.\hskip 1em plus 0.5em minus 0.4em\relax Berlin,
  Heidelberg: Springer-Verlag, 2011, pp. 576--587.

\bibitem{GMOT12}
M.~T. Goodrich, M.~Mitzenmacher, O.~Ohrimenko, and R.~Tamassia,
  ``Privacy-preserving group data access via stateless oblivious ram
  simulation,'' in \emph{Proceedings of the Twenty-third Annual ACM-SIAM
  Symposium on Discrete Algorithms}, ser. SODA '12.\hskip 1em plus 0.5em minus
  0.4em\relax SIAM, 2012, pp. 157--167.

\bibitem{Kush12}
E.~Kushilevitz, S.~Lu, and R.~Ostrovsky, ``On the (in)security of hash-based
  oblivious ram and a new balancing scheme,'' in \emph{Proceedings of the
  Twenty-third Annual ACM-SIAM Symposium on Discrete Algorithms}, ser. SODA
  '12.\hskip 1em plus 0.5em minus 0.4em\relax SIAM, 2012, pp. 143--156.

\bibitem{SSS12}
E.~Stefanov, E.~Shi, and D.~Song, ``Towards practical oblivious ram,'' in
  \emph{NDSS}, 2012.

\bibitem{SCSL11}
E.~Shi, T.-H.~H. Chan, E.~Stefanov, and M.~Li, ``Oblivious ram with o((logn)3)
  worst-case cost,'' in \emph{Proceedings of the 17th International Conference
  on The Theory and Application of Cryptology and Information Security}, ser.
  ASIACRYPT'11.\hskip 1em plus 0.5em minus 0.4em\relax Berlin, Heidelberg:
  Springer-Verlag, 2011, pp. 197--214.

\bibitem{GENTRY13}
C.~Gentry, K.~Goldman, S.~Halevi, C.~Julta, M.~Raykova, and D.~Wichs,
  ``\BIBforeignlanguage{English}{Optimizing oram and using it efficiently for
  secure computation},'' in \emph{\BIBforeignlanguage{English}{Privacy
  Enhancing Technologies}}, ser. Lecture Notes in Computer Science, vol.
  7981.\hskip 1em plus 0.5em minus 0.4em\relax Berlin, Heidelberg:
  Springer-Verlag, 2013, pp. 1--18.

\bibitem{PathORAM}
\BIBentryALTinterwordspacing
E.~Stefanov, M.~van Dijk, E.~Shi, C.~Fletcher, L.~Ren, X.~Yu, and S.~Devadas,
  ``Path oram: An extremely simple oblivious ram protocol,'' in
  \emph{Proceedings of the 2013 ACM SIGSAC Conference on Computer \&\#38;
  Communications Security}, ser. CCS '13.\hskip 1em plus 0.5em minus
  0.4em\relax New York, NY, USA: ACM, 2013, pp. 299--310. [Online]. Available:
  \url{http://doi.acm.org/10.1145/2508859.2516660}
\BIBentrySTDinterwordspacing

\bibitem{PrivateFS}
\BIBentryALTinterwordspacing
P.~Williams, R.~Sion, and A.~Tomescu, ``Privatefs: A parallel oblivious file
  system,'' in \emph{Proceedings of the 2012 ACM Conference on Computer and
  Communications Security}, ser. CCS '12.\hskip 1em plus 0.5em minus
  0.4em\relax New York, NY, USA: ACM, 2012, pp. 977--988. [Online]. Available:
  \url{http://doi.acm.org/10.1145/2382196.2382299}
\BIBentrySTDinterwordspacing

\bibitem{Shroud}
\BIBentryALTinterwordspacing
J.~R. Lorch, B.~Parno, J.~W. Mickens, M.~Raykova, and J.~Schiffman, ``Shroud:
  Ensuring private access to large-scale data in the data center,'' in
  \emph{Presented as part of the 11th USENIX Conference on File and Storage
  Technologies (FAST 13)}.\hskip 1em plus 0.5em minus 0.4em\relax San Jose, CA:
  USENIX, 2013, pp. 199--213. [Online]. Available:
  \url{https://www.usenix.org/conference/fast13/technical-sessions/presentation/lorch}
\BIBentrySTDinterwordspacing

\bibitem{oblivistore}
\BIBentryALTinterwordspacing
E.~Stefanov and E.~Shi, ``Oblivistore: High performance oblivious cloud
  storage,'' in \emph{Proceedings of the 2013 IEEE Symposium on Security and
  Privacy}, ser. SP '13.\hskip 1em plus 0.5em minus 0.4em\relax Washington, DC,
  USA: IEEE Computer Society, 2013, pp. 253--267. [Online]. Available:
  \url{http://dx.doi.org/10.1109/SP.2013.25}
\BIBentrySTDinterwordspacing

\bibitem{BurstORAM}
J.~Dautrich, E.~Stefanov, and E.~Shi, ``Burst oram: Minimizing oram response
  times for bursty access patterns,'' in \emph{USENIX Security}, 2014.

\bibitem{oram-isca13}
\BIBentryALTinterwordspacing
L.~Ren, X.~Yu, C.~W. Fletcher, M.~van Dijk, and S.~Devadas, ``Design space
  exploration and optimization of path oblivious ram in secure processors,'' in
  \emph{Proceedings of the 40th Annual International Symposium on Computer
  Architecture}, ser. ISCA '13.\hskip 1em plus 0.5em minus 0.4em\relax New
  York, NY, USA: ACM, 2013, pp. 571--582. [Online]. Available:
  \url{http://doi.acm.org/10.1145/2485922.2485971}
\BIBentrySTDinterwordspacing

\bibitem{roselli2000comparison}
D.~S. Roselli, J.~R. Lorch, T.~E. Anderson \emph{et~al.}, ``A comparison of
  file system workloads.'' in \emph{USENIX Annual Technical Conference, General
  Track}, 2000, pp. 41--54.

\bibitem{Yu14DSB}
X.~Yu, L.~Ren, C.~W. Fletcher, A.~Kwon, M.~van Dijk, and S.~Devadas,
  ``Enhancing oblivious ram performance using dynamic prefetching,'' Cryptology
  ePrint Archive, Report 2014/234, 2014, \url{http://eprint.iacr.org/}.

\end{thebibliography}

\end{document}